\documentstyle[12pt]{article}
\newlength{\extraspace}
\newlength{\extraspaces}
\newcommand{\be}{\begin{equation}
\addtolength{\abovedisplayskip}{\extraspaces}
\addtolength{\belowdisplayskip}{\extraspaces}
\addtolength{\abovedisplayshortskip}{\extraspace}
\addtolength{\belowdisplayshortskip}{\extraspace}}
\newcommand{\ee}{\end{equation}}

\newcommand{\bq}{\begin{eqnarray}
\addtolength{\abovedisplayskip}{\extraspaces}
\addtolength{\belowdisplayskip}{\extraspaces}
\addtolength{\abovedisplayshortskip}{\extraspace}
\addtolength{\belowdisplayshortskip}{\extraspace}}
\newcommand{\eq}{\end{eqnarray}}

\newcommand{\newsection}[1]{
\vspace{15mm}
\pagebreak[3]
\addtocounter{section}{1}
\setcounter{equation}{0}
\setcounter{subsection}{0}
\setcounter{footnote}{0}
\begin{flushleft}
{\large\bf \thesection. #1}
\end{flushleft}
\nopagebreak
\medskip
\nopagebreak}

\newcommand{\newsubsection}[1]{
\vspace{1cm}
\pagebreak[3]

\addtocounter{subsection}{1}
\noindent{ \bf \thesubsection. #1}
\nopagebreak
\vspace{2mm}
\nopagebreak}

\begin{document}
\addtolength{\baselineskip}{.8mm}

\thispagestyle{empty}
\begin{flushright}
{\sc PUPT}-1402\\
hep-ph@xxx/9305307 \\
May  1993
\end{flushright}
\vspace{.3cm}

\begin{center}
{\large\sc{Axions, Monopoles and Cosmic Strings}}

{\sc  Ian I. Kogan}
 \footnote{
on leave of absence from ITEP,
 B. Cheremyshkinskaya 25, Moscow, $117259$, Russia
}\\[2mm]
{\it Joseph Henry Laboratories\\[2mm]
Princeton University, Princeton, NJ 08544} \\[.3cm]

{\sc Abstract}
\end{center}
\noindent
We are discussing some aspects of the magnetic monopoles
 and cosmic strings interactions with  axion domain walss
 and membranes. The monopole  moving through  an axion domain
 wall is transformed into a monopole bag - the state with a
 dyon quantum number, but smaller mass.  In the case of
 an axion membrane the passing monopole  excites the
 chiral charged state at the membrane boundary. It will be
 shown that if  cosmic
 string  intersects  an  axion domain wall there
 will be $\theta = \pi$, i.e. maximal $CP$-violation inside
 the    string.   Strings carrying the flux of the $Z$-boson field
 become the sources of the baryon charge nonconseravtion.
 The symbiosis of the two pictures - monopole passing through an
 axion domain wall and string intersecting it - is the case
 of the interaction between an axion domain wall
 and a  Nambu electroweak string carrying
 $SU(2)_{L}$ monopoles at the ends.
\noindent
\newline
%%%%%%
\newline
%%%%%%%%%%
\newline
Talk given at the  International Workshop SUSY-93,
 Northeastern University, Boston, March 29 - April 1,  1993
\vfill

\newpage
\newsection{Introduction.}

\renewcommand{\footnotesize}{\small}

\noindent

 It is well known that QCD may have strong P and CP violation
 if the $\theta$ parameter, i.e. coefficient in front of topological
 charge $\frac{\theta}{32\pi^{2}}\tilde{F}_{a}^{\mu\nu}
F^{a}_{\mu\nu}$ is non-zero. There are several
  possible resolutions of this
 problem,
 but the most beatiful one is the idea of an axion \cite{axion}, i.e.
 the  light pseudoscalar particle $a(x)$ interacting with
 topological charge. One can get such a particle from the
 spontaneous breaking of the  global Peccei-Quinn (PQ)
  symmetry $U(1)_{PQ}$.
  Then axion appears as a phase of the order parameter of this
 symmetry  - the  scalar
 field $\Phi(x) = |\Phi| exp (ia)$.
 However the PQ  symmetry,  which can be  realised as
 a  shift  $a(x) \rightarrow a(x) + const$,
is explicitly  broken  by the  instanton effects in QCD
 and axion asquires a  mass by  order of
 $m_{a} \sim \Lambda_{QCD}^{2} /f_{a}$,
 where $f_{a} = <|\Phi|>$
is the scale of the spontaneous breaking of the PQ
 symmetry.

  There is another way to get an axion - from
 antisymmetric Kolb-Ramond \cite{kr}  field $B_{\mu\nu}$.
 Such field naturally
 arises in superstring theory \cite{gsw} and it was  shown
that the gauge-invariant field strength is defined as
\bq
H_{\mu\nu\lambda} = \partial_{[\mu}B_{\nu\lambda]}  -
 \Omega^{YM}_{\mu\nu\lambda} -
 \Omega^{G}_{\mu\nu\lambda}
\eq
where $\Omega^{YM}_{\mu\nu\lambda}$ and
 $\Omega^{G}_{\mu\nu\lambda}$
  are the three-forms dual to the Yang-Mills and
 gravitational topological charge densities. It is easy
 to see that equation of motion $~~
\partial_{\mu}H_{\mu\nu\lambda} = 0
{}~~$
 leads to
\bq
H_{\mu\nu\lambda} = \epsilon_{\mu\nu\lambda\sigma}\partial_{\sigma}\theta
\eq
However now $\partial^{2}\theta \neq 0$ because of the $\Omega$
 terms in $H_{\mu\nu\lambda}$ and  one gets
 $\partial^{2} \theta  =
Tr F \tilde{F} - Tr R\tilde{R}$
 (we shall not   consider  later the gravity contribution $Tr R\tilde{R}$),
  i.e. one gets axion coupling.
 What is important now - PQ symmetry $\theta \rightarrow
 \theta + const$ is really connected with a {\bf local} Kolb-Ramond
 symmetry $B_{\mu\nu} \rightarrow B_{\mu\nu} +
\partial_{[\mu}\epsilon_{\nu]}$.
 Indeed, we define not $\theta$, but $\partial_{\mu}\theta$ as dual to
 $H_{\mu\nu\lambda}$,
  thus
 by definition $\theta$ is defined up to an arbitrary constant.

In this talk we would like to discuss some old and new facts
 about interactions between some classical configurations
 of an axion field (axion strings and domain wals)
 and monopoles  and cosmic strings. A lot of interesting phenomena
 will appear due to the $\theta(x) F\tilde{F}$ term. The organization
 of the paper is as follows: in the next section we shall
 consider the interaction between magnetic monopoles and axion domain walls.
 Then we shall  discuss the monopole bag - the  charged monopole, i.e.
 the state with a dyon quantum numbers,
 in  a theory with an  axion  and finally the monopole
propagation through an axion membrane  will be considered.
In the section $3$ the
 interaction between cosmic string and axion domain wall will be
 considered  and appearence of the baryon charge nonconseravtion
  in this system will be discussed.

\newsection{ Monopole-Axion interaction}
\noindent

\newsubsection{ Axion domain wall}
\noindent

 Let us remember some facts about  the theory with monopoles in the presence
 of the $\theta$ term. The model  Lagrangian is
\bq
L = -\frac{1}{4}F_{a}^{\mu\nu}F^{a}_{\mu\nu}
 + \frac{e^{2}\theta}{32\pi^{2}}\tilde{F}_{a}^{\mu\nu}F^{a}_{\mu\nu}
 + L_{H}(\Phi)
\label{lagr}
\eq
where $e$ is the gauge coupling constant and
the last term is the scalar Lagrangian for the  Higgs field
 $\Phi^{a}$ in the adjoint representation
 of the gauge group.  It was shown by Witten
\cite{witten} that the monopole  electric charge   depends
 on $\theta$-angle
\bq
Q = e (\frac{\theta}{2\pi}) g + ne
\label{witten}
\eq
When $\theta$ is changed from $0$ to $2\pi$  one gets $n \rightarrow n + g$,
where $g$ is the magnetic charge (in our notation  minimal magnetic
charge is $1$,  $Q_{M} = 4\pi g/e$) and an electric
 change of  monopole  is changed.

It is easy to realise the process of the adiabatical
 variation of the vacuum angle $\theta$
  if one assumes that there is an  axion field  in the
 theory. Then $\theta$ parameter is not a number but a dynamical field -
 axion field $\theta(x)$  which is described by  the Lagrangian
\bq
L_{\theta} =\frac{ f_{\theta}^{2}}{2}
\partial_{\mu}\theta\partial_{\mu}\theta -
K^{2}(1-\cos\theta)
\label{Ltheta}
\eq
where parameter $K^{2}$ is connected with the nonperturbative part of the
 vacuum energy ($\theta$ dependence of the vacuum energy) and by order of
 magnitude is $\Lambda_{QCD}^{4} \approx (100 MeV)^{4}$.
 Thus the total lagrangian is the sum of the (\ref{lagr}) where
 $\theta$ angle is an axion field $\theta(x)$ and axion lagrangian
 (\ref{Ltheta}):
\bq
L = -\frac{1}{4}F_{a}^{\mu\nu}F^{a}_{\mu\nu}
 + \frac{e^{2}\theta(x)}{32\pi^{2}}\tilde{F}_{a}^{\mu\nu}F^{a}_{\mu\nu} +
L_{H}(\Phi) + L_{\theta}
\label{lagrtheta}
\eq
 There is an
  axion domain wall in this system  \cite{axiondomainwall}
  where the axion
 field depends only on distance $z$ to the wall surface -
$\theta(z) = 2\pi - 2\arccos \tanh (mz)$
 and the thickness of the wall  is of  order
   the inverse axion mass $m_{a} = K/f_{\theta}$,  which must be much
 smaller than
 the
monopole mass $M$. Then if the monopole adiabatically propagates  through
 the domain wall  one gets an apparent
 nonconservation of  an electric charge of the monopole. However the total
 electric charge must be conserved and
 as it have been  shown by Sikivie \cite{sikivie} it means that
 there must be an  induced electric charge on the wall.
 When the monopole
 is neutral this charge is  $+1/2$ and after the penetration
 the induced charge will be $-1/2$, so the total charge is conserved -
 $1/2 = 1 - 1/2$.

To find the induced electric charge on the domain wall it is necessary to
recall  why electric charge arises for nonzero $\theta$. As was
 shown in \cite{witten}, (see also \cite{gordon}, where a more general
 case including fermions was considered) the reason for $\theta$
 dependence of the monopole electric charge  is  because the
 appearence of  $\theta$-term  leads to the
 redefinition of the canonical momenta conjugate  to  gauge fields
$\vec{A}^{a}$
and it is easy to see  from (\ref{lagr}) that
 the new definition   still holds for  $x$-dependent
 axion field $\theta(x)$ \cite{kogan}
\bq
\vec{\Pi}^{a} = \frac{1}{i}\frac{\delta}{\delta \vec{A}^{a}} =
 -  \vec{E}^{a} + \frac{e^{2}\theta(x)}{8\pi^{2}}\vec{B}^{a}
\label{mom}
\eq
Let us work in the $A_{0}=0$ gauge and consider the   functional $|\Psi>$
which is an eigenstate of the magnetic and electric charges,
\bq
g|\Psi> = \frac{e}{4\pi}\oint \vec{dS} n^{a}\vec{B}^{a}|\Psi> \nonumber \\
Q|\Psi> = \oint \vec{dS} n^{a}\vec{E}^{a}|\Psi>
\eq
where $n^{a} = \Phi^{a}/|\Phi|$ and we use  factor $e/4\pi$ to normalize
 minimal magnetic charge $g=1$.  To
connect $g$ and $Q$  we must consider
 the Gauss law
\bq
\bigl[ i\vec{D}^{ab}\frac{\delta}{\delta \vec{A}^{b}} + J^{a}(\Phi)\bigr]
|\Psi> = 0
\eq
where $J^{a}(\Phi)$ is the scalar contribution, the exact form of which is
 irrelevant for us now. Using the expression for conjugate momenta
(\ref{mom}) one can see that because the Gauss law  must be correct
for arbitrary
$\theta$,  there is a relation between $g$ and $Q$:
\bq
Q|\Psi> = \frac{e^{2}}{8\pi^{2}}\oint \vec{dS} n^{a}\vec{E}^{a}|\Psi> =
(\frac{e}{2\pi})(\frac{e}{4\pi}) \oint \vec{dS}\theta(x)
n^{a}\vec{B}^{a}|\Psi>  \eq
Contrary to the constant $\theta$ case now we have to integrate the magnetic
  flux    weighted  with the
axion  field. Thus in the case of a domain wall we inegrate only
 on hemisphere where $\theta = 2\pi$ and  get $Q = e/2$ which is the
induced fractional
 charge of the domain wall in the presence of neutral monopole.
After monopole propagation through the wall and
   transmutation  into a  dyon with charge $e$
 the induced charge on the wall will be $-e/2$.

Let us note that  both the electric charge on the
 monopole and the induced electric charge on the domain wall
 are arising due the electric current corresponding to the
 $\theta(x) F\tilde{F}$ term in (\ref{lagrtheta}),
 which is not a total derivative for variable axion field
 $\theta(x)$.
  It is easy to see that one gets the
 conserved  electric  current  $J_{\mu}^{\theta} = J_{\mu}^{a}n^{a},~~
\partial^{\mu}J_{\mu}^{\theta} = 0$
 \bq
J_{\mu}^{\theta} = \frac{e^{2}}{16\pi^{2}}\partial_{\nu}
\left(\theta(x)n^{a}\tilde{F}^{a}_{\mu\nu}\right) =
\frac{e^{2}}{16\pi^{2}}\partial_{\nu}
\left(\theta(x)\tilde{F}_{\mu\nu}\right)
\nonumber \\
J_{0}^{\theta} = \frac{e^{2}}{8\pi^{2}}
 \vec{\bigtriangledown}(\theta(x) \vec{B}) =
\frac{e^{2}}{8\pi^{2}}\left(\theta(x) \vec{\bigtriangledown}\vec{B} +
  \vec{B}\vec{\bigtriangledown}\theta(x)\right) =
 \rho_{M} + \rho_{w} \label{rho} \\
\vec{J}^{\theta} = \frac{e^{2}}{8\pi^{2}}\left( \dot{\theta}\vec{B} +
 \vec{\bigtriangledown}\theta(x)\times\vec{E}\right)~~~~~~~~~
{}~~~~~~~~~~~~~~~~
\nonumber \\
 \eq
 where $F_{\mu\nu} = n^{a} F^{a}_{\mu\nu}$ is the electromagnetic field
and the current density is zero for constant $\theta$ as it must be,
 but the charge density has nontrivial contribution even at
 constant $\theta$ proportional to $\vec{\bigtriangledown}\vec{B}$,
 i.e. in the theories with the monopoles even the constant
 $\theta$ leads to new nontrivial effect - topological charge
 generates the electric charge in the presence of the
  magnetic monopole \cite{witten}.
 The charge density  $J_{0}$ is the sum of two terms
 $J_{0}^{\theta} =\rho_{M} + \rho_{w}$ which give the charge density
 in a monopole core with the charge  $Q_{M} =
 \int d^{3}x\rho_{M} = (\theta_{c}/2\pi)eg$, where
 $\theta_{c}$ is the axion field in the  core, and the charge
 density in the regions where the spatial derivatives of $\theta$  are
 non zero, i.e. precisely   inside the  domain wall.
  The induced charge on the domain wall
equals to $e/2$ or $-e/2$ depending on the position of the monopole in
 a complete agreement with the conservation of the total electric
 charge.  Let us mention also that condensed matter analogs of the
 axion domain walls were discussed in \cite{condmatteraxion}.

 The induced current $\vec{J}^{\theta}$ exists either when axion
 field depends on time  in the presence of magnetic field $\vec{B}$
or when it
 depends on coordinate
  (domain wall, for example) in the presence of electric
 field $\vec{E}$. Using the $\dot{\theta}B$ term one can easily get the
the  Witten result (\ref{witten})
 for the variation of the electric charge of the
monopole with respect to the variation of $\theta$ in time:
\bq
\Delta Q =
 \int dt \oint \,\vec{J}^{\theta}\vec{dS}  = \frac{e}{2\pi}  g \int dt
\dot{\theta}
 = \frac{e}{2\pi}  g \Delta\theta
\eq

Due to the oscillations of the axion field
 in the early Universe  the induced currents
  $\vec{J} \sim \dot{\theta}\vec{B}$ will arise
  in the regions with the  primordial magnetic
 fields. The damping of the axion field  oscillations
  will be  $\theta (t) \sim \theta_{0} \exp (im_{a}t)\exp[-\alpha^{2}
(B/f_{\theta})^{2} \sigma t]$ where $B$ is a typical value of the
 magnetic field and $\sigma$ is  some effective parameter
 which defines the loses of energy of the current
   $\vec{J}^{\theta}$  in the medium.
 However, because $B/f_{\theta}
 << 1$ it is unclear if this mechanism  leads to practically
 relevant damping  of  the primordial  axion oscillations.

\newsubsection{ Dyons and Monopole Bag}
\noindent

Let us note that  if one considers a closed domain wall one gets
   the integer
 induced charge on the wall if the monopole is inside ($1$ for minimal
 magnetic charge)  and $0$ if it is outside. It is amusing that the monopole
 can even stabilize the closed domain wall and the new stable state with
 the dyon quantum numbers $Q=g=1$  exists. It is into  this state, which was
  called  a {\bf monopole bag} in \cite{kogan}, the monopole will transform
 after penetrating through an axion domain wall.
  The reason why it can not
 be an ordinary dyon is very simple  -  dyon has both
 electric $\vec{E} \sim \vec{r}/r^{3}$ and magnetic $\vec{B} \sim
\vec{r}/r^{3}$ fields and the topological charge density
 $\vec{E}\vec{B} \sim 1/r^{4}$ is non zero.  We see from
 (\ref{lagrtheta}) that an  axion field
 interact with the topological
 density and the last one is a source for $\theta$.  That means that
   axion field $\theta(x) = 0$  is classically unstable. Let us consider
 the equations  of motion for the coupled axion and electromagnetic
 fields following  from the lagrangian (\ref{lagrtheta}).
 The  classical equations are:
\bq
f_{\theta}^{2}\partial^{2} \theta + K^{2}\sin\theta -
\frac{e^{2}}{8 \pi^{2}}\vec{E}\vec{B} = 0 \nonumber \\
\vec{\bigtriangledown}\vec{E} =
-\frac{e^{2}}{8 \pi^{2}}
\vec{\bigtriangledown}(\theta\vec{B}) + J^{0}   \\
\frac{\partial\vec{E}}{\partial t} +
\vec{\bigtriangledown}\times\vec{B} =
 - \frac{e^{2}}{8\pi^{2}}\left( \dot{\theta}\vec{B} +
 \vec{\bigtriangledown}\theta(x)\times\vec{E}\right) + \vec{J}
\nonumber
\label{system}
\eq
where  we used expressions   obtained for  axion  current $J^{\theta}$
in (\ref{rho}) and $J^{0},~\vec{J}$ are a charge
 density and a current corresponding to   all other  fields in
 the problem.
 If we  shall work outside the monopole core,  one
 can neglect the scalar fields and treats gauge fields as an
 abelian electromagnetic field. One can also neglect the dyon
 charge density $\rho_{D}$ which is non-zero only inside the monopole
core. Thus   all other currents exept $J^{\theta}$  are zero outside
the core.

 Let us assume that all fields depend only on time $t$ and radius $r$,
 then one can see that magnetic field does not depend on time and
 equals to monopole filed $\vec{B} =  (g/e) \vec{r}/r^{3}$ and
$\vec{\bigtriangledown}\times\vec{B} = 0$ as well as
 $\vec{\bigtriangledown}\theta \times \vec{E}=0$. Then from the last two
 equations one can find electric field:
\bq
\vec{E}(t,r) = -\frac{e^{2}}{8\pi^{2}}\theta(t,r) \vec{B}(r) +
\frac{e\vec{r}}{r^{3}}
\eq
where the first term is the axion contribution and the second  term
 is the dyon field. One can see that electric field has only radial
 component
\bq
E_{r}(t,r) =  \frac{e}{4\pi r^{2}}\left( 1 - g
  \frac{\theta(t,r)}{2\pi}\right)
\eq
Let us note that if $g>1$ the axion field can not screen the dyon charge
 completely  because at $r \rightarrow \infty$ one must have
 $\theta = 2\pi n$ - unless we have $N$ axion vacuua and $\theta$
 can takes the values $2\pi n/N$.  However  for physically interesting case
 $g = 1$ screening is possible. Later we shall consider only
 $ g = 1$. After all we get the selfconsistent equation for
 the axion field  $\theta(t,r)$:
\bq
f_{\theta}^{2}\ddot{\theta} - f_{\theta}^{2} \frac{1}{r^{2}}
\frac{\partial}{\partial r}\left(r^{2}
\frac{\partial \theta}{\partial r}\right)
+
K^{2}\sin\theta -
\frac{e^{2}}{32 \pi^{3}r^{4}}(1 - \frac{\theta}{2\pi}) = 0
\label{equation}
\eq
The monopole bag profile  is determined by  the stationary solution
of this equation $\theta(t,r) = \theta(r)$ with boundary conditions
$\theta(0) = 2\pi$ and $\theta(\infty) = 0$.
Before we shall discuss this solution let us notice
 that (\ref{equation}) is  the classical equation for the effective action
\bq
S_{\theta} = \int dt dr  r^{2}  \left[\frac{ f_{\theta}^{2}}{2}
(\dot\theta^{2} - (\partial_{r}\theta)^{2}) -
K^{2}(1-\cos\theta) - \frac{e^{2}}{32\pi^{2} r^{4}}
\left( \frac{\theta}{2\pi} - 1\right)^{2}\right]
\eq
where  the last term is nothing but the
 Coloumb energy
 $$\vec{E}^{2}/2 = (1/2)(e/4\pi r^{2})^{2}(1 -\theta/2\pi)^{2}.$$
The energy of the  stationary axion configuration
\bq
E = \int  dr  r^{2}  \left[\frac{ f_{\theta}^{2}}{2}
 (\partial_{r}\theta)^{2} +
K^{2}(1-\cos\theta) +  \frac{e^{2}}{32\pi^{2} r^{4}}
\left( \frac{\theta}{2\pi} - 1\right)^{2}\right]
\label{energy}
\eq
 is minimal on the stationary solution of (\ref{equation}).
 The  monopole bag mass is $M_{1} = M + E$.

 Let us
 note that the last two terms in the enrgy functional
 can be considered as some
 potential $V(\theta)$
\bq
V(\theta) = K^{2}(1-\cos\theta) +
  \frac{e^{2}}{32\pi^{2} r^{4}}
\left( \frac{\theta}{2\pi} - 1\right)^{2}
\eq
The behavior of this  potential is different at
 large and small $r$.
The local extrema of the potential are determined by the equation:
\bq
\frac{dV}{d\theta} = K^{2}\sin\theta + \frac{e^{2}}{32\pi^{3} r^{4}}
\left(\frac{\theta}{2\pi}-1\right) = 0
\eq
 which has three solutions -two minima and one maximum  in the interval
$ [0, 2\pi]$,
  (there may be other local extrema outside this interval) for
     $e^{2}/32\pi^{3}K^{2}r^{4}<<1$ and only one solution for
     $e^{2}/32\pi^{3} K^{2} r^{4} >> 1$.
 Thus  there is only one  minimum  at $\theta = 2\pi$ for  small $r$  but
  there are two minima - global at $\theta = 2\pi$ and local
 at $\theta$  in the interval $[0,\pi]$. The critical radius at which the
 local minimum disappears approximately equals to
 $r_{c} \approx (e^{2}/32\pi^{3} K^{2})^{1/4}$. As we shall see,  this
  radius is much
 larger then the bag radius $R_{B}$ and
  one can neglect the mass term $K^{2}\sin\theta$ in
 (\ref{equation})  at
 $ r < r_{c}$. The
 solution of (\ref{equation})  with
 $K=0$  is:
\bq
\theta(r) = 2\pi\left[1 - \exp\left(-\frac{e}{8\pi^{2}f_{\theta} r}
 \right)\right]
\label{solution}
\eq
 The bag radius now is $R_{0} = e/8\pi^{2} f_{\theta} << r_{c}$
and  we can use  solution (\ref{solution}) for all $r < r_{c}$. For
 $r > r_{c}$ one must take into account the mass term also, but in
 this region $\theta(r)$ is exponentially close to $0$ and one
 can neglects the  energy density in this region. To find the
 bag energy  we must substitute (\ref{solution}) into energy
 functional (\ref{energy}) and get \cite{kogan}
   $E = e f_{\theta}/ 8 \pi^{2}
 + o((K/f_{\theta}^{2})^{1/2}$ - this is the threshold energy
 for monopole to pass through the axion domain wall.
 We see that the
  scale $R_{0} = e/ 8 \pi^{2} f_{\theta}$ and the
 mass $M_{1} = M +  e f_{\theta}/ 8 \pi^{2}$ of the  monopole
 bag  practically do not depend on axion mass, i.e. on parameter $K$.

\newsubsection{ Axion membranes and axion strings.}
\noindent

We have considered  the case of an  infinite domain wall.
 More realistic is the  case  when  there is
  an axion membrane - a finite
 domain wall with a boundary, which is nothing but an axion string
\cite{axiondomainwall}, \cite{kls}.
 Let us remember that at high temperatures
 axion is massless (in the deconfining phase QCD instatntons
 do not lead to an axion mass)  and  there  are stable axion strings.
  At low temperatures $T < 100 MeV$, after
  the confinement-deconfinement transition,
  axion
 becomes massive and axion strings can not longer  exist.
 One can show that
  string-like distribution of the axion field will
 be transformed into finite domain walls, i.e. axion membranes,
   with the axion strings as
 boundaries  of the  membranes. Of course this membrane is unstable
 and will collapse, but it may takes some time and later we shall
  neglect  this instability.

  Let us take  a  monopole far from the membrane of the size $L$,
 then
  the total induced
 charge will be proportional to $L^{2}/R^{2}$, where $R$ is a
 distance between a  monopole and a membrane. For $R>>L$ it is
 small and if one starts with a monopole far from a membrane
 and ends with a monopole bag (after passing through the membrane)
 again far from the  membrane, i.e. in both cases the induced charges
 are zero ($R \rightarrow \infty$). It looks like that
 again we have some apparent nonconservation of the charge and
 the resolution of this paradox may be only one - now there  must  be
 induced charge
 on the boundary.  When monopole approaches the membrane
 ($R \sim L$)  there will be the
 induced charge $+1/2$ at the membrane and charge $-1/2$ at
 the boundary.
 After monopole will go through the membrane and take
 (as a monopole bag)
 the charge $+1$ with it, there are charges
  $-1/2$ on the membrane and $-1/2$
 on the boundary. Finally when monopole is  far from the membrane
 ($R >> L$) the charge $-1/2$ at the membrane will flow to the boundary
 and in the final state one has the total charge $-1$ at the boundary.
 Moreover, it will be not only induced charge, but also
 the induced current on the boundary - what we get is a realization
 of  a  Callan-Harvey effect \cite{ch} (see also  \cite{axionstring}).
The boundary of the membrane is an axion string which can carry chiral
 current - thus monopole passing through the  membrane will excite
 the  chiral currents at the boundary.

 It is easy to explain this phenomenon rewriting the inteaction between
 axion and topological charge as
 $(e^{2}/32\pi^{2})\int d^{4}x \partial_{\mu}\theta
 K_{\mu}$, where $K_{\mu}$ is the topological current (this is a
 one-form dual to the three form $\Omega$ in (3)).
  Taking into account that $\partial_{\mu}\theta \neq 0$ only
 inside  the membrane one gets  effective $2+1$ dimensional gauge
 theory with a Chern-Simons term  which are known to has  chiral gapless
 excitations on the boundary. This boundary ( or edge)
 excitations   are described by the
 chiral sector of some
 $1+1$-dimensional conformal field theory - the so called
 WZNW model.  Connection between
 $2+1$ Chern-Simons and $1+1$ conformal field theories was
  found in \cite{witten2}, for a review of applications
 to string theory and theory of edge excitations in the
 Quantum Hall effect see refs. \cite{carkogan} and \cite{wen}.

  This   charged current state at the boundary of the membrane will
 produce both electric $\vec{E}$ (because
 of the charge)  and magnetic $\vec{B}$ (because of the  current)
 fields
  and one can see that they carry non-zero angular momentum
 - this  can be considered as the  angular momentum
 of the axion membrane in the final state
 $\vec{M}_{a}
  = \int d^{3}x (\vec{E}\times\vec{B})\times\vec{r} =
 eg\vec{n}$, where $\vec{n}$ is the unit vector in the direction
 of the monopole motion. However the
 initial angular momentum  was zero and because
 the  total angular momentum must be conserved  there must be another
 contribution to the total angular momentum
  $\vec{M}$, which also includes the angular
 momentum  of the monopole-membrane system. The last one
  can be considered
 at large distances as two point particles -one of them (monopole bag)
 has both electric and magnetic charges, the second one (axion membrane)
 has only electric charge.  Remembering  the definition
 of the angular momentum operator for the charged particle in the
 presence of a magnetic  monopole
  $\vec{M}_{M} = \vec{L} - eg \vec{r}/|\vec{r}|$,
 where $\vec{L}$ is the usual orbital momentum and $\vec{r}$ is the
 radius vector between charge and monopole, one can see immediately that
 it is the last term
 $eg \vec{r}/|\vec{r}|$
 which also contributes to the final angular momentum
 and   precisely cancels the membrane contribution
 $\vec{M}_{a} - eg \vec{r}/|\vec{r}| = 0$.

\newsection{ Cosmic strings and axion domain walls}
\noindent

Let us  discuss now how the presence of  an  axion domain wall
   will  influence the  cosmic strings (for a review and
 references  about cosmic
 strings, see
 ref. \cite{vilenkin}).  Let us consider the
 simplest case of the Abrikosov-Nielsen-Olesen (ANO) string in the Abelian
 Higgs model with the   Lagrangian
\bq
L = -\frac{1}{4}F_{\mu\nu}F_{\mu\nu}
 + \frac{e^{2}\theta}{32\pi^{2}}\tilde{F}^{\mu\nu}F_{\mu\nu} +
 |D_{\mu}\Phi|^{2}
 - V(|\Phi|) + L_{\theta}
\label{u1}
\eq
where $V(|\Phi|)$ is the Higgs potential
 and $L_{\theta}$ is the axion lagrangian (5).
Neglecting the axion field one gets from (\ref{u1})
 the standard ANO string
 with nonzero $z$ component of
 the  magnetic field (if the string lies in the $z$-direction)
  $B(z,r) = m_{W}^{2} K_{0}(m_{W}r)$, where
 $m_{W}$ is the vector boson mass and $r$ are the radius
 in the transverse $x-y$ plane
 (we also assuming that Higgs mass $m_{H} >> m_{W}$ and
 are interesting in
 the distances $r >> m_{H}^{-1}$).

Now let us consider the same string in the presence of the
 axion domain wall
 and assume that string cross the wall in the transverse direction
 $z$. Then one has the induced charge
 inside
 the wall $\rho(z,r)  = -(e^{2}/8\pi^{2})
 \partial_{z} \theta(z)  B(r)$. One can easily
 writes the equation for the electric field (we are looking for the
 stationary solution, so $E_{i} = \partial_{i}A_{0}$ :
\bq
\partial^{2} A_{0} + m_{W}^{2} A_{0} = \rho(z,r)
\eq
 and  see that an  anzats $A_{0}(z,r) = \chi(z) B(r)$ is compatible with
 this equation and using the fact that
$\partial^{2} A_{0} = B(r) \partial_{z}^{2}\chi(z) +
 \chi(z) [r^{-1}\partial_{r}(r\partial_{r} B(r))]$ and $B$ obeys the
 equation (evrywhere outside the core of the string for $r> m^{-1}_{H}$)
 $r^{-1}\partial_{r}(r\partial_{r} B(r)) + m^{2}_{W}B(r) = 0$ one finally
 gets
\bq
\partial_{z}^{2}\chi =- \frac{e^{2}}{8 \pi^{2}} \partial_{z}\theta (z); ~~~
E_{z} = \partial_{z}A_{0} =
  -\frac{e^{2}}{8 \pi^{2}} B(r)[\theta(z) + const]
\eq

The choice of constant is dictated by the boundary conditions on
 $\theta(\pm \infty)$. If there is no domain wall, $const =0$,
 in the case of the domain wall when $\theta (-\infty) = 0$ and
 $\theta(+\infty) = 2\pi$ one has $const = -\pi$.
 However if we have both electric and magnetic fields we again have the
 source for an  axion field - as it was in the case of a dyon - and
 the equation for the axion field now are:
\bq
f_{\theta}^{2}\partial^{2} \theta + K^{2}\sin\theta -
\frac{e^{2}}{8 \pi^{2}}\vec{E}\vec{B} = \nonumber \\
f_{\theta}^{2}\partial^{2} \theta + K^{2}\sin\theta +
(\frac{e^{2}}{8 \pi^{2}})^{2}B^{2} (\theta - \pi) = 0
\eq
The effective potential for axion field   is
$V_{eff}(\theta) = -K^{2}\cos\theta + (e^{4}/128 \pi^{4})B^{2}(\theta-
\pi)^{2}$
 Inside the string (i.e. at $r < m_{W}^{-1})$)
 there is only one minimum at $\theta =\pi$ - and only at large
 $r$ where $e^{2}B(r) \sim \Lambda_{QCD}$ the usual minima at
 $\theta = 0$ and $2\pi$ will  appear. This means that in the region
 occupied by  a flux of the gauge field one can not get a
 domain wall structure at all - he  has  $\theta = \pi$
  for all $z$  and at only at large  $r >  m_{W}^{-1}$
 there will be  restoration of the usual domain wall structure
 $\theta(-\infty) = 0$ and $\theta(+\infty) = 2\pi$.
 The real situation even more complicated because one must take into
 account the influence of the $A_{0}$ on the $|\Phi|$, which
 leads to $z$-dependence of the vector boson mass $m_{W}$.
 However the qualitative result that inside the cosmic string
 one will get $\theta = \pi$ is correct.
 One can say that the interior of the domain wall  occupies
 the cosmic string.  Because $\theta = \pi$  electric
 field
 $E_{z} = 0$ inside the flux region. However at
  the transition region,  i.e. at the boundary of
 the flux  $r\sim m_{W}^{-1}$) we shall get $z$ dependence
 and in result there is  nonzero $F\tilde F = B E_{z} =
  -(e^{2}/8 \pi^{2})B^{2} [\theta(z) - \pi]$.

The same qualitative picture will be correct and for strings
 carrying the  flux of the $Z$-boson field,  which can arise in the
 standard electroweak theory (\cite{nambu},\cite{semilocal}).
In this case nonzero $\vec{E}\vec{B}$ leads to the violation of the
 baryon charge \cite{th} $\partial_{\mu}J^{\mu}_{B} \sim Z_{\mu\nu}
\tilde{Z}_{\mu\nu}$. It is interesting to note that because $\theta
 = \pi$ inside the string we have at the same time the  maximal
 possible  strong
 CP-violation.
  Considereing a  long closed string intersecting the domain wall
  and being   divided into   two parts
 (in general nonequal)
   one can get the production of the baryon charge (the sign will
 depend on what part of the closed string  has larger length).

 It is also interesting to consider the case of the Nambu
 \cite{nambu}  electroweak string carrying $SU(2)$ monopole-antimonopole
 pair at the ends. The isospinor Higgs field
 $\Phi^{i},~i = 1,2$ near one of the ends takes the form (in polar coordinates)
\bq
\Phi^{i} = |\Phi| \left( \begin{array}{c}
 \cos \frac{1}{2}\phi \\
 \sin \frac{1}{2}\phi\, e^{i\psi}
\end{array}\right)
\eq
which is ill-defined at negative $z$ axis  $\phi = \pi$ and in this region
 one has string - at the second end of which there is an antimonopole.
 Far away from the system there is a linear combination
 of $SU(2)_{L}$ and $U(1)_{Y}$ gauge fields, i.e. precisely
 electromagnetic  $U(1)_{em}$, created by the monopole-antimonopole
 pair connected by the string. Along the string the $U(1)_{Y}$
 part has a return flux, whereas the $SU(2)_{L}$ part does not
  and the poles are genuine $SU(2)$ monopoles.
  After passing through the domain
 wall these  monopoles  will carry $SU(2)$ electric charges  -
 now it will be monopole-antimonopole bags.
 Again  one   has  non-zero $F\tilde{F}$ - but in this case outside
 the domain wall. One can stabilize this string making  it
 rotating - sending such an object through an axion domain
 wall one will get after some stable system carrying nontrivial
 topological charge density.  The detailed description of
 this process and the properties of the charged  rotating strings
  are  unknown - for example it is not clear how the rotation
 of the electroweak Nambu string will affect the monopole
  transformation into a monopole bag. It is also unclear
    what will be
 the production rate for the baryon charge
    and can these objects  decay into smaller strings
 by producing monopole-antimonopole pair in the middle of
 the string.

\newsection{ Conclusion }
\noindent

We  discussed here the different aspects of the interaction between
 axion domain walls and membranes and monopoles and cosmic strings.
 The appearence of the induced charges in the presence of the
  external gauge fields leads to a very interesting physics.
 Especially interesting is the problem of the
 cosmic string interaction with  an axion domain wall. The new
 string-domain wall configuration arising during the intersection
 with the maximal strong $CP$-violation ($\theta = \pi$) inside the
 string,  which at the same time is a source for baryon number
 nonconservation, is very interesting. The case of the
 Namby $SU(2)$ string with monopole-antimonopole pair at the ends
 is an amusing example of a symbiosis of the two process
 we had considered - monopole propagation throught
 an axion domain wall  and infinite string intersecting it.
 It will be interesting to understand the  behaviour
 of the stable spinning $SU(2)$ string, especially to understand
 the details of the $SU(2)$  monopole transformation into monopole bag
  in the presence of the string  rotation.   The detialed picture
 of the anomalous production of the baryon charge  is also of
 greate interest. These and many other intersting physical effects
 which will arise in these systems are
  definitely  deserve further investigations.

{\bf Acknowledgements.}

\noindent

\bigskip

It is a pleasure to thank Martin Bucher, Curtis Callan, Renata Kallosh,
   Larry  McLerran, Pierre  Sikivie, Neil Turok, Tanmay Vachaspati and
 Frank Wilczek
  for interesting  discussions. This work was supported by the
 National Science Foundation grant NSF PHY90-21984.

{\renewcommand{\Large}{\normalsize}

\end{document}